\documentstyle[12pt,epsf]{article}
\begin{document}

\hyphenation{mono-pole mono-poles lat-tice}
\newcommand{\w}{\left<W(R,T)\right>}
\newcommand{\wmon}{W_{mon}^e}
\newcommand{\wph}{W_{ph}^e}
\newcommand{\ts}{\textstyle}

\begin{titlepage}

\begin{tabbing}
\` arch-ive/9809019 \\
\` ILL-(TH)-98-\#14 \\
\` August, l998 \\
\end{tabbing}
 
 
\begin{center}
{\bf Abelian Links, Monopoles, and Glueballs\\
in $SU(2)$ Lattice Gauge Theory \\}
\vspace*{.5in}
John D. Stack and Rafal Filipczyk\\
\vspace*{.2in}
{\it Department of Physics \\
University of Illinois at Urbana-Champaign \\
1110 W. Green Street \\
Urbana, IL 61801 \\}
\vspace*{0.3in}
{\it }
\end{center}

\begin{tabbing}
\` PACS Indices: 11.15.Ha\\
\` 12.38.Gc\\
\end{tabbing}

\end{titlepage}
\vfill\eject

\pagestyle{empty}

\begin{center}
{\bf Abstract}
\end{center}

\noindent   We investigate the masses of $0^{+}$ and $2^{+}$ glueballs
in $SU(2)$ lattice
gauge theory using abelian projection to the maximum abelian gauge.
We calculate glueball masses using both abelian links and monopole
operators.  Both methods reproduce the known full $SU(2)$ results
quantitatively.  Positivity problems present in the abelian projection
are discussed.  We study the dependence of the glueball masses on magnetic
current loop size, and find that the $0^{+}$ state requires a much greater
range of sizes than does the $2^{+}$ state.

\vspace*{.5in}
 
\newpage
\pagestyle{plain}

\section{Introduction }
\label{sec-intro}

The attempt to understand confinement in QCD has continued ever since
Gell-Mann and Zweig introduced the quark concept.  Attention has 
focussed almost universally 
on the pure gauge theory
without light dynamical quarks as the place to begin study of the
non-perturbative physics involved in confinement.  
In the absence of light dynamical
quarks, what is meant by confinement can be precisely defined.
Namely, in a confining theory, Wilson's loop must show an area law,
or what is equivalent, the fundamental string tension must be finite.

Having simplified full QCD to pure $SU(3)$ gauge theory, many theorists have
gone one step further and simplified to pure $SU(2)$ gauge theory.
The physics of confinement is thought to be quite similar for
any of the $SU(N)$ groups, of which the simplest is $SU(2)$.  While an
$SU(2)$ gauge theory obviously cannot explain the spectrum of baryons,
lattice gauge theory simulations have shown that
the meson spectra of pure $SU(2)$ and $SU(3)$ gauge theories are
remarkably alike.

A major effort has been devoted to the theoretical understanding of
confinement in $SU(2)$ pure gauge theory, both in the continuum and on
the lattice.  Within this framework, the dominant theme has been to
search for a topological explanation of confinement.  Topological objects
are postulated to dominate the Euclidean path integral and lead to
an area law for Wilson loops.  The topological objects usually considered
are instantons, monopoles, and vortices.  Of these, only instantons are firmly
established semi-classically 
in Euclidean pure $SU(2)$ gauge theory.  However, the
general consensus of work done on instantons is that while they are surely
highly relevant for understanding the breakdown of chiral symmetry in
QCD, they appear to have little to do with confinement.

This leaves monopoles and/or vortices as possible agents of confinement in
pure $SU(2)$ gauge theory.  The present work is devoted to further exploring
the monopole approach.  Since the work of 't Hooft and 
Polyakov \cite{thoopolymon},  monopoles
are well-known to exist semiclassically if the gauge theory 
contains adjoint scalar fields in addition to the usual gauge fields.
Such scalar fields are natural in N=2 supersymmetric
$SU(2)$ gauge theory, and the famous work of Witten and Seiberg showed
that non-perturbative phenomena in this theory are indeed driven by
the 't Hooft-Polyakov monopoles present in the theory \cite{witten}.  
However, the
relevance of the
Witten-Seiberg work for understanding confinement in pure $SU(2)$ gauge
theory remains uncertain.  

The suggestion that monopoles nevertheless exist and control
confinement in the pure non-abelian gauge theory
was made much
earlier
by 't Hooft~\cite{thoo}, who also advocated partial gauge-fixing as a way to
see monopole physics more clearly.  The basic physical picture of 
confinement via monopoles is the ``dual superconductor", where the linear
potential between heavy quark and anti-quark originates in a electric
flux tube surrounded by the magnetic current of a monopole condensate.
The fundamental string tension has been successfully 
calculated in this framework 
\cite{jssnrw,bali}.
The purpose of the present work is to see if the glueball
spectrum can also be explained.  Glueballs are as
characteristic of a confining theory as the string tension, and any approach
which purports to explain confinement should  also produce quantitative 
results for the glueball spectrum.  

To make the paper self-contained, the abelian projection formalism is
reviewed in the next Section.  The reader familiar with this can go immediately
to Section \ref{sec-gballs}. We will work in lattice units throughout,
and since all states are even under charge conjugation in $SU(2)$ we will
denote glueballs by their spin $J$ and parity $P$ as $J^{P}$.

\section{~The Abelian Projection}
\label{sec-abel}
In 't Hooft's framework,
the first step is a partial gauge-fixing, applied only to those gauge fields
which are ``charged", or have off-diagonal generators in the Lie algebra of
the gauge group.  
 The central idea is that the
monopoles associated with the abelian gauge invariance left after
partial gauge-fixing will control
non-perturbative phenomena.  
For an $SU(2)$ gauge group with the generator $T_3$ diagonal, the
gauge field $A^{3}_{\mu}$ is the abelian field or ``photon", and gauge-fixing
is done only on $A^{a}_{\mu}, a=1,2$, or equivalently the charged fields
\begin{displaymath}
W^{\pm}_{\mu} = \frac{1}{\sqrt{2}}(A^{1}_{\mu} \pm iA^{2}_{\mu}) .
\end{displaymath}
After gauge-fixing, the charged field are ignored, the important physics
being postulated to reside solely in the abelian gauge field and its monopoles.

The 't Hooft approach has been explored rather extensively 
in $SU(2)$ lattice gauge theory \cite{polikarpov}.
The maximum abelian gauge  has been found to be the only
form of the gauge-fixing condition which  allows a quantitative 
abelian calculation of the
string tension.
In the continuum form of the maximum
abelian gauge, the continuum functional
\begin{equation}
G_{c}\equiv \frac{1}{2}\sum_{\mu}\int \left ( (A^{1}_{\mu})^{2}
+(A^{2}_{\mu})^{2} \right )d^{4}x
=\sum_{\mu}\int(W^{+}_{\mu}W^{-}_{\mu})d^{4}x
\label{contgf}
\end{equation}
is minimized over all $SU(2)$ gauge transformations, leading to the 
conditions
\begin{displaymath}
(\partial_{\mu}+igA^{3}_{\mu})W^{+}_{\mu}=
(\partial_{\mu}-igA^{3}_{\mu})W^{-}_{\mu}=0,
\end{displaymath}
where $g$ is the $SU(2)$ gauge coupling.

Stated precisely the purpose of the present paper is to see if the glueball
spectrum can also be explained in this gauge.

\subsection{~Lattice Gauge-Fixing}~The $SU(2)$ lattice gauge theory is built out of link variables
$U_{\mu}(x)$,
\begin{displaymath}
U_{\mu}(x)\equiv e^{iga\vec{A}_{\mu}\cdot\vec{\tau}},
\end{displaymath}
where $\vec{\tau}=\vec{\sigma}/2$ are the generators of $SU(2)$ in the 
fundamental representation, and $a$ is the lattice spacing.  On the lattice,
the maximum abelian gauge is obtained by maximizing the lattice functional
\begin{equation}
G_{l}=\sum_{x,\mu}\frac{{\rm tr}}{2}\left[U^{\dagger}_{\mu}(x)\sigma_{3}U_{\mu}(x)
\sigma_{3}\right]
\label{lattgf}
\end{equation}
over all $SU(2)$ gauge transformations~\cite{kron}.  It is easy to show that 
in the continuum limit,
maximizing $G_{l}$ is equivalent to
minimizing $G_{c}$.

The demand that $G_{l}$ be stationary with respect to a gauge transformation
at  an arbitrary site $y$ leads to the requirement that
\begin{equation}
X(y)\equiv \sum_{\mu}\left[U_{\mu}(y)\sigma_{3}U^{\dagger}_{\mu}(y)
+U^{\dagger}_{\mu}(y-\hat{\mu})\sigma_{3}U(y-\hat{\mu})\right]
\label{Xdef}
\end{equation}
be diagonal.  This can be accomplished by a gauge transformation
$\Omega(y)$.  However, the value of $X$ at the nearest neighbors of $y$ is
affected by $\Omega(y)$, so the diagonalization of $X$ over the whole lattice
must be done iteratively.

It is useful to factor a $U(1)$ link variable from $U_{\mu}(x)$, writing 
$
U_{\mu}(x)=u_{\mu}(x)w_{\mu}(x),
$
where 
$u_{\mu}(x)=\exp(i\phi_{\mu}^{3}\tau_{3})$, and
$w_{\mu}(x)=\exp(i\vec{\theta_{\mu}}\cdot\vec{\tau}),$
with $\theta^{3}_{\mu}\equiv0$.
The angle $\phi^{3}_{\mu}$ can be extracted 
from the matrix elements of $U_{\mu}$
by expanding the gauge-fixed $SU(2)$
link $U_{\mu}$ in Pauli matrices, writing
\begin{equation}
U_{\mu}=U_{\mu}^{0}+i\sum_{k=1}^{3}U_{\mu}^{k}\cdot \sigma_{k}.
\end{equation}
Then $\phi^{3}_{\mu}=2\cdot\arctan(U_{\mu}^{3}/U_{\mu}^{0})$.
The  $SU(2)$ action is periodic in $\phi^{3}_{\mu}$ with period
$4\pi$.  To carry over standard formulas from $U(1)$ lattice gauge
theory, it is more convenient to use an angle  
 $\bar{\phi}^{3}_{\mu} \equiv \phi^{3}_{\mu}/2$, with
 $\bar{\phi}^{3}_{\mu} 
\in (-\pi,\pi]$.

\subsection{~Monopole Location in SU(2)}
The location of the magnetic current  starts with plaquette
angles $\bar{\phi}^{3}_{\mu\nu}$ constructed from
$\bar{\phi}^{3}_{\mu}$,
\begin{equation}
\bar{\phi}^{3}_{\mu\nu}(x)=
\partial_{\mu}\bar{\phi}^{3}_{\nu}
-\partial_{\nu}\bar{\phi}^{3}_{\mu}=
\bar{\phi}^{3}_{\mu}(x)
+\bar{\phi}^{3}_{\nu}(x+\hat{\mu})-\bar{\phi}^{3}_{\mu}(x+\hat{\nu})
-\bar{\phi}^{3}_{\nu}(x)
\end{equation}
The plaquette angle $\bar{\phi}^{3}_{\mu\nu}$ is resolved into a 
Dirac string contribution, plus a fluctuating part:
\begin{equation}
\bar{\phi}^{3}_{\mu\nu}=2\pi \bar{n}_{\mu\nu}+\tilde{\phi}^{3}_{\mu\nu},
\label{eqnphimunu}
\end{equation}
where 
$\tilde{\phi}^{3}_{\mu\nu}\in (-\pi,\pi]$ 
and $\bar{n}_{\mu\nu}$ is an integer~\cite{degrand}.
The integer-valued magnetic current $\bar{m}_{\mu}$ is determined by 
the net flux
of Dirac strings into an elementary cube:
\begin{equation}
\bar{m}_{\mu}
=-\frac{1}{2}\epsilon_{\mu\nu\alpha\beta}
\cdot\partial_{\nu} \bar{n}_{\alpha\beta}.
\label{eqnnmunu}
\end{equation}

In the 
identification of $\bar{m}_{\mu}$, elementary cubes were used, so
a magnetic charge is located at the center of a spacial 1-cube,
likewise for other components of the magnetic current.  This is done
mainly for practical reasons; if (say) 2-cubes 
were used instead, the effective lattice size would become $8^4$ instead of
$16^4$.  
\subsection{~Abelian Wilson Loops  } 
The maximum abelian gauge has in a global way made the charged link angles
$\vec{\theta_{\mu}}$ as small as possible. Non-perturbative quantities are
postulated to be solely contained in the abelian link angles
$ \phi^{3}_{\mu}$ or equivalently $\bar{\phi}^{3}_{\mu}$.  This allows two
abelian
approximations to the full $SU(2)$ Wilson loop.  In the first, we simply
use the link angles $\bar{\phi}^{3}_{\mu}(x)$ in the standard $U(1)$
formula for a Wilson loop:
\begin{equation}
 W_{U(1)} =
\left< 
\exp \left(i\sum_{x}\bar{\phi}^{3}_{\mu}J_{\mu}\right)\right>,
\label{u1basic}
\end{equation}
where $\left< \cdot \right>$ denotes the expectation value over
the ensemble of $\bar{\phi}^{3}$ link angle configurations, and 
$J_{\mu}$ is an integer-valued current defined by the path of the
heavy quark.

In $U(1)$ lattice gauge theory, a  Wilson 
loop expressed as in Eq.(\ref{u1basic})
can be factored into a
term arising from photon exchange, times a 
term arising from monopoles,
\begin{equation}
 W_{U(1)}  =  W_{phot}  \cdot
 W_{mon} .
\label{u1fac}
\end{equation}

The photon contribution is perturbative, so for non-perturbative 
quantities, the same information should be in $W_{mon}$ as $W_{U(1)}$.
The monopole Wilson loop $ W_{mon} $ forms our second
abelian approximation to the full $SU(2)$ Wilson loop.
An explicit formula for $ W_{mon} $ 
is obtained by writing
$J_{\mu}$ as the curl of
a Dirac sheet variable~\cite{diracsh};
$J_{\mu}=\partial_{\nu}D_{\mu\nu}$, 
where $\partial_{\nu}$ denotes a discrete derivative.
Then $ W_{mon} $ is given by
\begin{equation}
W_{mon} =
                \left<\exp\left(\frac{i2\pi}{2}\sum_{x}
D_{\mu\nu}(x)
                   F^{*}_{\mu\nu}(x)\right)\right>_m ,
\label{eqnmon}
\end{equation}
where
$\left<\cdot\right>_m$ denotes the sum over configurations of magnetic current.
\footnote{ There is in principle a finite volume correction to 
Eq.(\ref{eqnmon})  arising from Dirac sheets, \cite{stack_vol}, 
but we have checked that it is negligible in the present calculations.}
For a rectangular Wilson loop, a useful choice 
for the sheet variable $D_{\mu\nu}$ is to set
$D_{\mu\nu}=1$ on the
plaquettes of the flat rectangle
with boundary  $J_{\mu}$, and $D_{\mu\nu}=0$ on all other 
plaquettes.  
In Eq.(\ref{eqnmon}), $F^{*}_{\mu\nu}$ is the dual of the field strength
due to the magnetic current;
\mbox{$F^{*}_{\mu\nu}(x)=\frac{1}{2}\epsilon_{\mu\nu\alpha\beta}F_{\alpha\beta}(x)$.}
The field strength itself is derived from a magnetic vector potential 
$A^{m}_{\mu}$,
\mbox{$F_{\mu\nu}\equiv\partial_{\mu}A^{m}_{\nu}-\partial_{\nu}A^{m}_{\mu}$,} 
where
\begin{equation} A^{m}_\mu(x)=
\sum_{y}v(x-y)\bar{m}_{\mu}(y),
\label{eqnamag}
\end{equation}
and $\bar{m}_{\mu}$ is the integer-valued, conserved magnetic current
defined in Eq.(\ref{eqnnmunu}).   

To summarize, once we have gone to 
the maximum abelian gauge, our two abelian approximations to $SU(2)$
Wilson loops are Eq.(\ref{u1basic}) or Eq.(\ref{eqnmon}).  If the basic
assumptions involved in the abelian projection are correct, non-perturbative
quantities like the string tension and glueball spectrum should be 
obtainable from either one.

\section{Glueball operators }
\label{sec-gballs}

	We will make use of standard methods for extracting the glueball
spectrum in lattice gauge theory, as expounded in particular by Teper
and collaborators ~\cite{ishikawa}.  The basic object of interest is
$\Phi$, called the glueball ``wave function".  This is a gauge-invariant
combination of $SU(2)$ links constructed so as to make 
$|\langle\Omega | \Phi | G\rangle|$ as large as possible, where $|\Omega>$ is the 
$SU(2)$ vacuum, and $ |G>$ is a one glueball state, usually taken to
have a definite three-momentum.  In practice, $\Phi$ is a relatively
simple Wilson loop-like combination of $SU(2)$ links, summed over
orientations to project a definite spin-parity, and over spacial locations
to project a definite three-momentum.  An important advance made by Teper
was to notice that $|<\Omega | \Phi | G>|$ is numerically much larger if
$\Phi$ is constructed out of ``fuzzy" links \cite{teper_fuzz}. 
Fuzzy links are constructed by
replacing original  products of two links in a given direction 
by themselves plus some fraction of
``staples" (four-link products with the same endpoints as the original link).
A related procedure applied directly to single links was introduced by the
Ape collaboration \cite{ape}.  This is commonly called ``smearing", and is
what we use in the present work, when calculating with abelian links.
To keep everything strictly local in Euclidean time, the staples used in
smearing are spacial, {\it i.e.} for a $z-$link, the smeared staples are
$x-z-x$ and $y-z-y$, but not $t-z-t$.  Smearing is typically done several
times in the full gauge theory \cite{ape}, but here we smear the abelian
links  once, twice, or three times.

For the $0^{+}$ and $2^{+}$ glueballs, $\Phi$ is constructed out of 
simple rectangular Wilson loops, anchored at the spacial point $\vec{x}$,
and lying in a particular spacial plane.
The spin projection for $0^{+}$ is done by averaging over the three 
spacial planes,
$x-y$, $x-z$, and $y-z$, while for the
$2^{+}$ ,
the spin projection is accomplished 
by  subtracting twice the $x-y$ rectangular loop from the sum of $z-x$ and
$z-y$ loops \cite{ishikawa}.
Projection to a three-momentum $\vec{p}=0$ is done
by summing over spacial points $\vec{x}$, divided by $N_{x}N_{y}N_{z}$, where
$N_{x}, N_{y},$ and $N_{z}$ are the number of lattice sites in x,y, and 
z-directions.

\subsection{~Glueballs and the Abelian Projection}
In the present work, we evaluate the Wilson loops in the wave functions $\Phi$
either in terms of the abelian link angles $\bar{\phi}^{3}$
as in Eq.(\ref{u1basic}),
or in terms of the magnetic current of monopoles as in Eq.(\ref{eqnmon}).  Using these,
we calculate correlation functions which we label as
$$C(t)_{U(1)}=\langle\Phi(t)\Phi(0)
\rangle_{U(1)}/\langle\Phi(0)\Phi(0)\rangle_{U(1)}$$ 
$$C(t)_{mon}=\langle\Phi(t)\Phi(0)
\rangle_{mon}/\langle\Phi(0)\Phi(0)\rangle_{mon},$$
where t refers to the separation between
the two time-slices where $\Phi$ is evaluated.  Any vacuum expected value of
$\Phi$ is assumed to have been removed.  
These calculations are analogous to calculating
the potential using Eq.(\ref{u1basic}) and Eq.(\ref{eqnmon}), respectively.  
The aim 
of course is to see if the glueball spectrum can be obtained in the maximum
abelian gauge to a similar degree of accuracy as the string tension.  The use of smeared links
is straightforward in the calculation which uses abelian links directly, namely
$C(t)_{U(1)}$, and for this case we will report on our results both with and without smeared abelian
links.
The Wilson loops in the monopole correlation function are obtained by computing the dual
flux through the surface of the loop.  Calculating this flux through loops with smeared edges
is somewhat complicated, so for the monopole correlation function
we simply calculate directly using the simplest plane rectangular surface  .  
In any case, after gauge-fixing to the maximum abelian gauge, 
smearing is not essential
for obtaining good results, unlike the situation for full $SU(2)$.  
The process of going to the maximum abelian gauge and using either
the resulting abelian links or monopoles results in correlation functions with much less noise
than the full $SU(2)$ correlation functions.  

\subsection{~Positivity}
  The full $SU(2)$ correlation function
$C(t)_{SU(2)}$ is of the form
\begin{equation}
C(t)_{SU(2)} =  \Sigma_{i}c_{i}\cdot e^{-E_{i}t}
\label{eqnsu2fit}
\end{equation}
where we again assume any vacuum expected value of $\Phi$ has been removed,
and for simplicity treat the case of a system finite in space
but infinite in  time. Since $C(t)_{SU(2)}$ is normalized to 1.0 at $t=0$,
we have
$$c_{i}=|\langle\Omega|\Phi|i
\rangle|^{2}/\sum_{i}|\langle\Omega|\Phi|i\rangle|^{2},$$
so 
the non-zero coefficients $c_{i}$ are positive and certainly less than unity. 
On a  lattice with finite time extent and periodic boundary conditions,
we must make the replacement 
$e^{-E_{i}t} \rightarrow e^{-E_{i}t} + e^{-E_{i}(T-t)}$ in Eq.(\ref{eqnsu2fit})
to satisfy periodicity.  In addition, a constant $b$, independent of $t$ but
of order $\exp{-T}$ will in general be present in the correlation function
\cite{schierholz}.  
It is easy to show that for an $SU(2)$ action 
with nearest neighbor couplings in
time and a positive transfer matrix, the coefficients 
$c_{i}$ continue
to be positive and $<1.0$.  Our calculations were done with the Wilson
$SU(2)$ action, which satisfies the above properties.

The correlation function $C(t)_{U(1)}$ is obtained by replacing each
$SU(2)$ link in $\Phi$ by its $U(1)$ approximation in 
the maximum abelian gauge.  
The correlation function $C(t)_{mon}$ involves a further approximation
in which only the contributions of monopoles are retained, and photons
or abelian gluons are dropped.

The correlation functions $C(t)_{U(1)}$ and
$C(t)_{mon}$
are fitted to a form \cite{schierholz}
\begin{equation}
 b +c_{0}\cdot(e^{-E_{0}t}+e^{-E_{0}(T-t)}).
 \label{fit}
 \end{equation}
The fit is  aimed at getting
only the lowest energy eigenvalue, $E_{0}$, in a given channel.
The results  show that the
constant $b$ is always very small, essentially zero, while
the value of of $E_{0}$ 
is generally in agreement 
within statistical errors
with the corresponding full $SU(2)$ value.  

Problems with positivity show up in two ways.  The first has to do with the
constant $c_{0}$.
In the full $SU(2)$ correlation function
a good choice of the operator $\Phi$ will generally increase the value of
$c_{0}$, but positivity always requires $c_{0}<1.0$.  
However in $C(t)_{mon}$ and $C(t)_{U(1)}$, fits of the form of Eq.(\ref{fit}) 
do result in $c_{0}>1.0$. The problem is more severe for $C(t)_{mon}$
where values of $c_{0}>1.3$ are common.  

The second difficulty is that an acceptable fit of the form Eq.(\ref{fit}) 
is often
not possible without dropping small t points, $t=0$,
$t=0,1$, or in some extreme cases $t=0,1,2$.  This could happen
for a function which does satisfy positivity, 
if the lowest eigenvalue $E_{0}$ is
not sufficiently dominant.  In such a case, the 'effective eigenvalues'
$$ E_{0}(t) \equiv -\ln(C(t+1)/C(t))$$
will {\it decrease} with increasing $t$, reflecting the fact that the exact
correlation function of
the form Eq.(\ref{eqnsu2fit}), is concave upward.  The failure of the fit to a simple
exponential would in this case merely represent the 
importance of physical excited states in the correlation function.
However for $C(t)_{U(1)}$ and $C(t)_{mon}$, it
is often true that the values of $E_{0}(t)$ ${\it increase}$ with $t$ for $t=0,1$, showing
that $C(t)_{U(1)}$ and $C(t)_{mon}$ violate this concavity requirement for small
values of $t$.   
Our strategy for dealing with this problem violation is 
simple.  We examine the effective eigenvalues $E_{0}(t)$ 
and look for a 'plateau'.
If for small values of t, the $E_{0}(t)$ are {\it less} than
those in the plateau, those values of t are omitted in the fit.
The fit then runs
from some $t_{min}$ to t=7 on our $16^4$ lattice.  This procedure is of course
heuristic, and is based on the assumption that for sufficiently large values of
t, the true lowest eigenvalue of the underlying 
$SU(2)$ gauge theory will dominate
in $C(t)_{U(1)}$ and $C(t)_{mon}$.

In a given channel, we use rectangular Wilson loops
of various sizes as operators, and compute only diagonal terms, {\it i.e.} 
correlations of the same operator with
itself.  In the resulting fits, we favor results where the value of $c_{0}$ 
is as close to 1.0 as possible,
that is we disfavor small $c_{0}$ since it implies that the operator has a small
overlap with the state of interest, and we also disfavor values of $c_{0}$ much
larger than 1.0 since they imply a large amount of positivity violation.  Detailed
examples will be given in the next section, where it will be seen that despite these
problems with positivity, the final results are surprisingly reliable.

It is perhaps worth making some general comments as to why positivity violation occurs
at all.  Since we have a gauge system, a configuration of link angles on a given
time-slice represents a physical configuration plus many redundant
variables.  One of the reasons for gauge-fixing is to reduce the number of irrelevant
variables.  However, as opposed to a gauge which acts on some 
local operator, putting a link angle 
configuration into the maximum abelian gauge
is an iterative process, 
involving many sweeps of the lattice, and therefore coupling sites which are far apart
on the lattice.
While the non-locality
thereby introduced must cancel out if only $SU(2)$ gauge-invariant operators are used,
this is not true when links are represented only by their $U(1)$ parts.  
The effective action which generates $U(1)$ configurations in the maximum abelian
gauge no longer has the properties of the original Wilson $SU(2)$ action, in particular,
it will not be nearest-neighbor in time. Correlation
functions can then violate standard positivity properties.

\section{Calculations}
\label{sec-calcns}
\subsection{Simulation and Gauge-Fixing}

	Our simulations were done on a $16^4$ lattice, using the standard
Wilson form of the $SU(2)$ action, at  
$\beta=2.40,$ and $2.50$.  At each $\beta$, after equilibration, $2000$
configurations were saved.  Saved configurations were separated by $20$ updates
of the lattice, where a 
lattice update consisted of one heatbath sweep~\cite{creutz},
plus one or two overrelaxation sweeps~\cite{overr}. 
Each of these configurations was
then projected into the maximum abelian gauge using the overrelaxation 
method of
Mandula and Ogilvie, with their parameter $\omega=1.70$~\cite{mandula}. 
The overrelaxation process was stopped after the off-diagonal elements
of $X(x)$ of Eq.(\ref{Xdef}) were sufficiently small.  Expanding $X(x)$
in Pauli matrices,
\begin{equation}
X=X^{0}(x)+i\sum_{k=1}^{3}X^{k}(x)\cdot \sigma_{k},
\end{equation}
we used
\begin{equation}
\left<|X^{ch}|^{2}\right>\equiv \frac{1}{ L^{4}}\sum_{x}\left(
|X^{1}(x)|^{2}+|X^{2}(x)|^{2}\right)
\end{equation}
as a measure of the average size of the off-diagonal matrix elements of 
X over the lattice,
and required $\left<|X^{ch}|^{2}\right>\le 10^{-10}$.  This condition was reached 
in approximately 1000 overrelaxation sweeps.  Even though our condition on
 $\left<|X^{ch}|^{2}\right>$ is very tight, the gauge-fixed configuration is
 not unique; Gribov copies can still occur.  
 However, Hart and Teper have shown that the uncertainty in the string tension
 caused by the Gribov ambiguity is rather small \cite{hart_gribov}.
 We have assumed the same will hold for the glueball spectrum.


From each gauge-fixed $SU(2)$ link, the $U(1)$ 
link angle $\bar{\phi^{3}_{\mu}}$ was extracted using the formula
$\bar{\phi}^{3}_{\mu} =\arctan(U^{3}_{\mu}/U^{0}_{\mu})$,  
as described in   Section~\ref{sec-intro}.  Then making use of the plaquette angles
$\bar{\phi}^{3}_{\mu\nu}$,
the magnetic current
$\bar{m}_{\mu}(x)$ was found.  

\subsection{The Glueball Spectrum in Maximum Abelian Gauge }
\label{sub-gmax}

In this section, we present our results for the glueball spectrum 
in detail.  Our fits
are all of the form Eq.(\ref{fit}).  In Tables 1-8 below we present the
data from our fits for the $0^{+}$ and $2^{+}$ channels at zero momentum for 
for abelian links and monopoles at $\beta=2.50$, and $\beta=2.40$.
Since the three-momentum is zero, $E_{0}$ is in fact
an estimate of the mass of the corresponding glueball , denoted as $M_{0}$ in the
tables.  The column labelled
'operator' labels the Wilson loop used in  calculating $C(t)_{U(1)}$
or $C(t)_{mon}$.
Thus in Table 1, the operator $2 \times 1$ means the correlation function
involves a $2 \times 1$ Wilson loop correlated with itself, while
the operator $(2 \times 1)_{-1}$ means a $2 \times 1$ loop smeared once, etc.
The column labelled 'fit' lists the small t points omitted, if any,
in the fit. 
The next column  gives the fitted value
of the coefficient $c_{0}$ in the fit, which we henceforth call the
'overlap'.  

We have also studied (less extensively) the glueballs masses as derived from
correlation functions with finite momentum.  We took the momentum to be
the lowest possible on the lattice,
$\pi/8$ in lattice units, in (say) the $x$-direction.  The value of $E_{0}$ was
derived from the corresponding correlation function in the same manner as for
zero momentum.  Then the mass $M_{0}$ was obtained by assuming  the usual
(continuum) relation between $E_{0}$ and $M_{0}$, 
$$ E_{0} = \sqrt{ M_{0}^{2} + p^{2}}.$$
Representative finite momentum results at $\beta=2.50$ are presented in Table 9.

Before presenting our final estimates of glueball masses, we first remark
on some systematic trends which can be seen in the various tables.  For
data obtained from  
the link correlation functions (Tables 1-4), the overlap coefficient
$c_{0}$ is safely within error bars of the physical range $ 0 < c_{0} <1$
for almost all cases.  In contrast, for the monopole correlation functions
(Tables 5-8), $   c_{0} >1$ is the rule rather than the exception.
Here we only find $ c_{0} <1$ for the first few operators for the
$2^{+}$ state at $\beta=2.50$.  So for the link correlation functions,
this gross violation of positivity is rather uncommon,  while it is 
quite common for the data obtained from monopole correlation functions.
Nevertheless, 
the values of $E_{0}$ given in the tables for the two cases indicate that
link and monopole operators are coupling to the same set of states.

The need to drop small values of $t$ from the fits to
Eq.(\ref{fit}) is another indication of positivity violation, caused in
each case by small values of the effective eigenvalues $E_{0}(t)$.  This
dropping of some points is always necessary when $   c_{0} >1$, the extreme
case being the $2^{+}$  at $\beta=2.5$ state from monopoles, where 
$ c_{0} >1.4$ required dropping $t=0,1,2$.  However, for $ c_{0} <1$,
low effective eigenvalues for small $t$ also necessitate dropping points
at small $t$.  The $2^{+}$ at $\beta=2.5$ state from monopoles also
illustrates this.

We now turn in Table \ref{tab-su2} to a comparison of our results 
for glueball masses with 
the best available full $SU(2)$ data.  The latter is from Michael and
Teper \cite{michael-teper}  who extracted glueball masses from 
20,000  measurements at $\beta=2.40$ on a $16^4$ lattice, 
and 14,000 at $\beta=2.50$ 
on a $20^4$ lattice.  Generally the agreement is good between monopoles
and links and with the full $SU(2)$ results, especially for
the lightest and heaviest states, namely the $0^{+}$ at $\beta=2.50$
and the $2^{+}$ at $\beta=2.40$.  For the remaining states, there are
some discrepancies at roughly the 5\% level.  This is not unreasonable,
given the heaviness of glueballs vs. the square root of the string tension,
and the relatively small number of measurements (2000) used in the present
work.

\subsection{~Magnetic Current Loop Size and the Glueball Spectrum}

In the maximum abelian gauge,  the magnetic current can be resolved into
closed loops containing different numbers of links.  
At $\beta=2.50$ , the percentage
of magnetic current in loops containing less than 10, 20, 50 and 100 links is
43\%, 51\%, 56\%, and 59\%, respectively \cite{jssnrw}.  It has been known
for some time that the fundamental string tension is unaffected by the small
loops of magnetic current.  At $\beta=2.50$, dropping all current contained
in loops with {\it less} than 100 links does not affect the string tension.
On the other hand, if we increase the size of the smallest allowed
magnetic current
loop to 200 links, 
the resulting string tension is approximately 10\%
smaller than the correct value.  So we can say that loops $\sim 100$ links
and larger are necessary to explain the $\beta=2.50$ string tension.  

In this
section, we explore the dependence of the glueball spectrum on magnetic
current loop size, discussing  for simplicity only $\beta=2.50$.  For
the $0^{+}$ state, we will use $2\times 1$ Wilson loops in  $C(t)_{mon}$,
for the $2^{+}$ state, we will use $2 \times 2$ Wilson loops.  The
first question we asked is whether the masses are stable under the 
elimination of small magnetic current loops.  
We tested this by successively eliminating all
loops with 
{\it less}  than 10, 20, 50, 100 {\it etc.} links.  If the glueball masses were to
behave in the same manner as the string tension, the masses would be
preserved up to a  lower limit on sizes of $\sim 100$ links, but then would fall
below their correct values.  In fact the glueball masses behave somewhat
differently than the string tension.  For the $0^{+}$ state, using the entire
magnetic current, and the $2\times 1$ Wilson loop in $C(t)_{mon}$,
we obtained a mass of 0.72(2).  While we obtain values consistent with this
when we eliminate all magnetic current loops smaller than 6, 10, and 20 links, by the time
we eliminate all loops containing less than 50 links ,
the $0^{+}$ mass has fallen to 0.60(3) , and steadily 
declines as we increase the
size of the smallest loop of magnetic current allowed.  Thus the $0^{+}$
glueball state is sensitive to smaller loops of magnetic current than
the string tension.

For the $2^{+}$
state, using the entire magnetic current and the $2\times 2$ Wilson loop in
$C(t)_{mon}$, we obtained a mass of 0.98(8).  We find that this mass is 
stable in successively eliminating loops of size less than 10,\ 20,\ 
50,\ 100 links , and even somewhat beyond.  (For the lower limit of 100 links,
we obtained a mass of 0.98(2).)
So the lower bound on sizes of loops of magnetic
current needed to reproduce the $2^{+}$ glueball is quite similar to that needed for
the string tension. 

Since glueball masses, like the string tension, are non-perturbative quantities,
it might seem obvious that they would require the largest loops of magnetic
current.
 The presence of these very large loops of magnetic current is thought
to be a defining characteristic of a confining phase.  For example, in
$U(1)$ lattice gauge theory, they are present in the confining phase, and
absent in the deconfined Coulombic phase \cite{js7-2}.  To investigate
this in the $SU(2)$ glueball spectrum, we
tried extracting the  masses from $C(t)_{mon}$ after eliminating all 
loops of magnetic 
current {\it larger} than a certain
size.  We find that the $0^{+}$ and $2^{+}$ states behave rather differently under these
cuts.  For the $0^{+}$, if the {\it upper} limit on loop sizes is taken
to be 100 links, the resulting mass is close to the value obtained with
the entire magnetic current, 0.75(3), {\it vs.} 0.72(2).  However, upon
raising the upper limit to 200 links, the mass is too small , 0.53(5) , 
and remains
so when the upper limit on loop size is raised still further giving
0.63(5) when the upper limit is 1000 links.  Only when the upper limit
is raised greater than 1000 links does the mass finally stabilize, reaching
0.70(8) by the time the largest loop of magnetic current contains 1500
links.  Thus the $0^{+}$ state is sensitive to the largest loops of magnetic
current.

For the $2^{+}$, the story is different.  When the upper limit on loop size
is 100 links, the mass value obtained is too large, 1.48(6), {\it vs.}
0.98(8) obtained with the entire magnetic current.  But by the time 
the upper limit is
raised to 200 links, the mass is 1.05(15), and remains within error bars
of 0.98(8) as the upper limit is raised steadily to 2000 links.  So unlike
the $0^{+}$ and the string tension, explaining the 
$2^{+}$ mass does not appear to require the largest loops of magnetic current.

In Table \ref{tab-cuts}, we show $0^{+}$ and $2^{+}$ mass at 
$\beta=2.50$ with various cuts on the number of links allowed in loops
of magnetic current.  In the Table, $N_{min}$ denotes the number of links
in the smallest allowed magnetic current loop, and $N_{max}$ does likewise
for the largest.  There are two rather intriguing results.  The first is
the $0^{+}$ state basically requires all of the magnetic current which
resides in loops $\ge$ 50 links in size. (Note that this still means that
over 50\% of the current can be
discarded.)  The second is the relative insensitivity
of the $2^{+}$ state to the particular cut made on magnetic current.
The $2^{+}$ state is visible in a wide variety of windows, the exception being
one which
only allows small loops.

An interesting difference between $0^{+}$ and $2^{+}$ states  
has also been found in the 
work of 
Sch\"{a}fer and Shurak \cite{shuryak} on instantons and glueballs.  
We now discuss the possible relation of their results  to our own.
In the instanton case, the $0^{+}$ channel receives
strong non-perturbative contributions from individual instantons and
anti-instantons, as well as instanton-anti-instanton interactions.
The $2^{+}$ channel, on the other hand, does not receive contributions from
individual instantons and anti-instantons, only a rather weak contribution
from instanton-anti-instanton interactions.  
On the lattice, 
it is known that a distribution of instantons and anti-instantons,
when cast into the maximum abelian gauge,
is 
represented by a network of magnetic current \cite{hart}, and
a rough correspondence can be made between the size of an instanton
and the size of the magnetic
current loop  it generates. Since instantons, anti-instantons and their
interactions all give a non-perturbative contribution
to the $0^{+}$ correlation function, 
but only the interactions  contribute to 
the $2^{+}$, it is natural for the $0^{+}$ state to 
depend more strongly on magnetic current loops of all sizes than the $2^{+}$.  
While this conclusion is similar in the two approaches, our
work would imply that  the $0^{+}$ still 
could not be quantitatively 
explained in an instanton gas/liquid, 
since the latter,
having no confinement \cite{chen}, 
contains too few of the
very largest loops of magnetic current.

\section{Conclusions and Summary}
\label{sec-concl}

We have shown that contact can be made with the spectrum of  glueballs in
$SU(2)$ lattice gauge theory, using the maximum abelian gauge and computing
correlation functions using either
abelian links or monopoles.  The calculations we have presented show
rather convincingly that for sufficiently large values of $t$,
both
$C(t)_{U(1)}$ and $C(t)_{mon}$ are dominated by the lightest glueball 
in the channel of interest.  The present calculations were done starting
with 2,000 $SU(2)$ configurations, as opposed to the $O(20,000)$ used
in conventional $SU(2)$ calculations.  In addition, when glueballs are
found using $C_{mon}(t)$, only the magnetic current is needed, which 
occupies around 1\% (at $\beta=2.50$) of the links on 
the lattice \cite{jssnrw}.  So despite the effort
required to project a configuration into the maximum abelian gauge, 
the description finally involves a rather small subset of
the variables in a full $SU(2)$
configuration.   Further, the results on how different components of the
magnetic current build up the $0^{+}$ and $2^{+}$ give  information
not readily obtainable in a conventional $SU(2)$ calculation.  

Nevertheless, the present method of calculation;
gauge-fixing, followed by a truncation of the degrees
of freedom, is not a controlled approximation, and the results must  
be checked by comparing them with
those from
full $SU(2)$ calculations.  One of the major uncertainties
which shows up in the glueball calculations reported here
is the presence of positivity
violation.  In calculations of the heavy quark potential, this was hardly
visible at all.  The only sign of it was a very small wrong sign Coulomb
term seen in the potential calculated from monopoles \cite{jssnrw}.  No
positivity violation was seen in the potential calculated with abelian links.
In contrast, the glueball calculations done here show positivity violation 
occasionally with
abelian links, and quite commonly with monopoles.  
It would clearly be of great interest to
know the spectrum of masses 
which are allowed to appear in 
$C(t)_{U(1)}$ and $C(t)_{mon}$. If only physical states can occur, then 
positivity violation would imply that
excited states  may have coefficients with negative signs.  \footnote{
This happens in 
conventional calculations which use smearing and a 
cold-wall source \cite{ape}.}
In this situation, 
standard upper bound statements on glueball masses are lost,
but the spectrum still contains only physical states.
In the maximum abelian gauge, 
the greatly reduced statistical errors of $C(t)_{U(1)}$ and $C(t)_{mon}$
vs $C(t)_{SU(2)}$  might make this a price worth paying.

We are interested in extending the present work in several directions.
The first goal is to include the $0^{-}$ and $2^{-}$, the next states up 
in mass in the
$SU(2)$ glueball spectrum.  Beyond this, we want to gather a much larger
dataset to pin down any possible discrepancies between abelian link/monopole
glueball masses  and those obtained with full $SU(2)$, and finally 
attempt to extract excited state
masses and compare those with full $SU(2)$ .  This could shed
light on the allowed spectrum of $C(t)_{U(1)}$ and $C(t)_{mon}$, discussed 
above.  The explanation of the glueball spectrum is clearly a challenge
that any topological approach to confinement must meet.

This work was supported in part by the National Science Foundation under
Grant No. NSF PHY 94-12556.  During the early stages of this work,
R.\ F.\ was supported in part by Department of Energy grant 
DE-FG02-91ER40677.  The calculations were carried out on the Cray
 T-90 system at the San Diego Supercomputer Center at the University of 
 California, San Diego, and the Power Challenge
 system at the National Center for Supercomputing Applications at
University of Illinois,  both supported in part by the National Science
Foundation.


%
%

\clearpage

\begin{center}
\section*{Tables}
\end{center}

\begin{table}[h]
\caption{ $0^{+}$ state from abelian links at $\beta=2.50$}
\vspace{0.5cm}
\begin{center}
\begin{tabular}{||c|c|c|c||} \hline
    operator & fit & $c_{0}$ & $M_{0}$ \\ \hline
    $2\times 1$ & 0 & 0.81(2) &   0.76(2) \\ \hline
    $(2\times 1)_{-1}$ &   & 0.99(1) &   0.68(1) \\ \hline
    $2\times 2$ & 0 & 0.91(2) &   0.70(2) \\ \hline
    $(2\times 2)_{-1}$ & 0,1 & 1.15(8) &   0.68(4) \\ \hline
    $3\times 2$ & 0 & 0.92(2) &   0.67(2) \\ \hline
    $3\times 3$ & 0,1 & 1.04(7) &   0.67(4) \\ \hline
    $4\times 4$ & 0,1 & 1.04(7) &   0.63(4) \\ \hline
\end{tabular} 
\end{center}
\end{table}

\begin{table}[h]
\caption{ $2^{+}$ state from abelian links at $\beta=2.50$}
\begin{center}
\begin{tabular}{||c|c|c|c||} \hline
    operator & fit & $c_{0}$ & $M_{0}$ \\ \hline
    $3\times 3$ & 0 & 0.63(2) &   1.17(5) \\ \hline
    $(3\times 3)_{-1}$ & 0 & 0.95(3) &   1.06(3) \\ \hline
    $(3\times 3)_{-2}$ &   & 0.992(5) &   1.04(1) \\ \hline
    $4\times 4$ & 0 & 0.62(6) &   1.02(4) \\ \hline
    $(4\times 4)_{-1}$ & 0 & 1.05(3) &   0.99(3) \\ \hline
    $5\times 5$ & 0 & 0.55(2) &   0.94(4) \\ \hline
\end{tabular} 
\end{center}
\end{table}

\begin{table}
\caption{ $0^{+}$ state from abelian links at $\beta=2.40$}
\begin{center}
\begin{tabular}{||c|c|c|c||} \hline
    operator & fit & $c_{0}$ & $M_{0}$ \\ \hline
    $1\times 1$ & 0 & 0.93(3) &   1.10(4) \\ \hline
    $(1\times 1)_{-1}$ &   & 0.94(3) &   0.99(3) \\ \hline
    $2\times 1$ & 0 & 0.99(3) &   1.07(3) \\ \hline
    $2\times 2$ & 0,1 & 1.2(2) &   1.04(7) \\ \hline
\end{tabular} 
\end{center}
\end{table}

\begin{table}
\caption{ $2^{+}$ state from abelian links at $\beta=2.40$}
\begin{center}
\begin{tabular}{||c|c|c|c||} \hline
    operator & fit & $c_{0}$ & $M_{0}$ \\ \hline
    $2\times 2$ & 0 & 0.67(9) &   1.7(1) \\ \hline
    $(2\times 2)_{-1}$ & 0 & 0.86(8) &   1.5(1) \\ \hline
    $3\times 3$ & 0,1 & 0.73(6) &   1.48(9) \\ \hline
    $(3\times 1)_{-1}$ & 0 & 0.7(1) &   1.6(1) \\ \hline
    $(4\times 2)_{-1}$ &   & 1.00(1) &   1.45(2) \\ \hline
\end{tabular} 
\end{center}
\end{table}

\begin{table}
\caption{ $0^{+}$ state from monopoles at $\beta=2.50$}
\begin{center}
\begin{tabular}{||c|c|c|c||} \hline
    operator & fit & $c_{0}$ & $M_{0}$ \\ \hline
    $2\times 1$ & 0 & 1.19(2) &   0.72(2) \\ \hline
    $2\times 2$ & 0,1 & 1.38(9) &   0.70(4) \\ \hline
    $3\times 1$ & 0 & 1.19(2) &   0.69(2) \\ \hline
    $3\times 3$ & 0,1,2 & 1.9(4) &   0.72(8) \\ \hline
    $4\times 4$ & 0,1,2 & 2.0(4) &   0.67(7) \\ \hline
\end{tabular} 
\end{center}
\end{table}

\begin{table}
\caption{ $2^{+}$ state from monopoles at $\beta=2.50$}
\begin{center}
\begin{tabular}{||c|c|c|c||} \hline
    operator & fit & $c_{0}$ & $M_{0}$ \\ \hline
    $2\times 2$ & 0,1 & 0.9(1) &   0.98(8) \\ \hline
    $3\times 1$ & 0,1 & 0.6(1) &   1.0(1) \\ \hline
    $4\times 1$ & 0,1 & 0.7(1) &   0.93(8) \\ \hline
    $5\times 1$ & 0,1 & 0.7(1) &   0.94(9) \\ \hline
    $3\times 2$ & 0 & 1.21(3) &   0.95(2) \\ \hline
    $3\times 3$ & 0,1,2 & 1.4(5) &   0.9(1) \\ \hline
    $4\times 4$ & 0,1,2 & 1.8(6) &   0.9(1) \\ \hline
\end{tabular} 
\end{center}
\end{table}

\begin{table}
\caption{ $0^{+}$ state from monopoles at $\beta=2.40$}
\begin{center}
\begin{tabular}{||c|c|c|c||} \hline
    operator & fit & $c_{0}$ & $M_{0}$ \\ \hline
    $1\times 1$ & 0 & 1.33(4) &   1.04(2) \\ \hline
    $2\times 1$ & 0 & 1.34(3) &   0.98(2) \\ \hline
    $2\times 2$ & 0,1 & 1.6(1) &   0.97(5) \\ \hline
    $3\times 1$ & 0 & 1.34(3) &   0.94(2) \\ \hline
\end{tabular} 
\end{center}
\end{table}

\begin{table}
\caption{ $2^{+}$ state from monopoles at $\beta=2.40$}
\begin{center}
\begin{tabular}{||c|c|c|c||} \hline
    operator & fit & $c_{0}$ & $M_{0}$ \\ \hline
    $2\times 2$ & 0 & 1.39(7) &   1.39(6) \\ \hline
    $3\times 1$ & 0 & 1.35(7) &   1.53(5) \\ \hline
    $4\times 1$ & 0 & 1.25(6) &   1.47(5) \\ \hline
    $5\times 1$ & 0 & 1.23(8) &   1.48(7) \\ \hline
    $6\times 1$ & 0 & 1.17(8) &   1.47(7) \\ \hline
\end{tabular} 
\end{center}
\end{table}    

\begin{table}
\caption{ Finite momentum results from links and monopoles at $\beta=2.50$}
\begin{center}
\begin{tabular}{||c|c|c|c|c||} \hline
    state     & type      & operator    & $c_{0}$ & $M_{0}$ \\ \hline
    $0^{+}$  & links     & $2\times 1$ & 0.81(2) & 0.74(3) \\ \hline
    $2^{+}$  & links     & $4\times 4$ & 0.66(3) & 1.06(5)  \\ \hline
    $0^{+}$  & monopoles & $2\times 1$ & 1.21(2) & 0.66(2)  \\ \hline
    $2^{+}$  & monopoles & $2\times 2$ & 1.24(4) & 1.07(3)  \\ \hline
\end{tabular} 
\end{center}
\end{table}

\begin{table}
\caption{ Glueball masses from full $SU(2)$, links, and monopoles}
\label{tab-su2}
\begin{center}
\begin{tabular}{||c|c|c|c|c||} \hline
    $\beta$ & state & $SU(2)$ & links & monopoles\\ \hline
    $2.5$ & $0^{+}$ & 0.72(3) & 0.69(1) & 0.70(1)  \\ \hline
    $2.5$ & $2^{+}$ & 1.05(3) & 1.04(1) & 0.95(2)  \\ \hline
    $2.4$ & $0^{+}$ & 0.94(3) & 1.06(2) & 0.99(1)  \\ \hline
    $2.4$ & $2^{+}$ & 1.52(3) & 1.47(2) & 1.49(4)  \\ \hline
\end{tabular} 
\end{center}
\end{table}

\begin{table}
\caption{ The $0^{+}$ and $2^{+}$ masses with cuts on the magnetic current}
\label{tab-cuts}
\begin{center}
\begin{tabular}{||c|c|c|c||} \hline
    $N_{min}$ & $N_{max}$ & $0^{+}$ & $2^{+}$  \\ \hline
         0    & $\infty$  &  0.72(2) & 0.98(8)   \\ \hline
         20   & $\infty$  &  0.69(4) & 1.00(2)   \\ \hline
         50   & $\infty$  &  0.61(3) & 0.97(2)   \\ \hline
         100  & $\infty$  &  0.49(2) & 0.98(2)   \\ \hline
         20   & 2000      &  0.69(5) & 0.98(2)   \\ \hline
         20   & 1000      &  0.51(5) & 1.00(3)   \\ \hline
         50   & 2000      &  0.61(5) & 0.98(2)   \\ \hline
         100  & 1000      &  0.40(6) & 0.96(3)   \\ \hline
         100  &  500      &  0.1(1)  & 0.95(4)   \\ \hline
\end{tabular} 
\end{center}
\end{table}    

\newpage

\begin{center}
\section*{ Figure Captions}
\end{center}

\vspace{.15in}

\noindent {\bf Figure 1}.  
 The $0^{+}$ correlation function
 from abelian links. Symbols: \\
 $diamonds-\ (2\times 1)_{1}$ loops, $crosses-\ 2\times 2$ loops ,
 $squares-\  3\times 2$ loops.

\vspace{.15in}

\noindent {\bf Figure 2}.  
 The $0^{+}$ correlation function
 from monopoles Symbols: \\
 $diamonds-\ (2\times 1)_{1}$  loops,  $crosses-\ 2\times 2$ loops ,
 $squares-\  3\times 1$ loops.

\vspace{.15in}

\noindent {\bf Figure 3}.  
        The $2^{+}$ correlation function
	from abelian links.  Symbols: \\
	$diamonds-\ (3\times 3)_{1}$  loops,  
	$crosses-\ (4\times 4)_{1}$ loops ,
	$squares-\  4\times 4$ loops.

\vspace{.15in}

\noindent {\bf Figure 4}.  
        The $2^{+}$ correlation function
	from monopoles. Symbols: \\
	$diamonds-\ 2\times 2$  loops,
	$crosses-\ 3\times 2$ loops ,
	$squares-\  3\times 1$ loops.

\newpage

\newpage

\epsfverbosetrue
        \begin{figure}
        \begin{center}
        \leavevmode
        \epsfxsize=6.0in
        \epsfbox{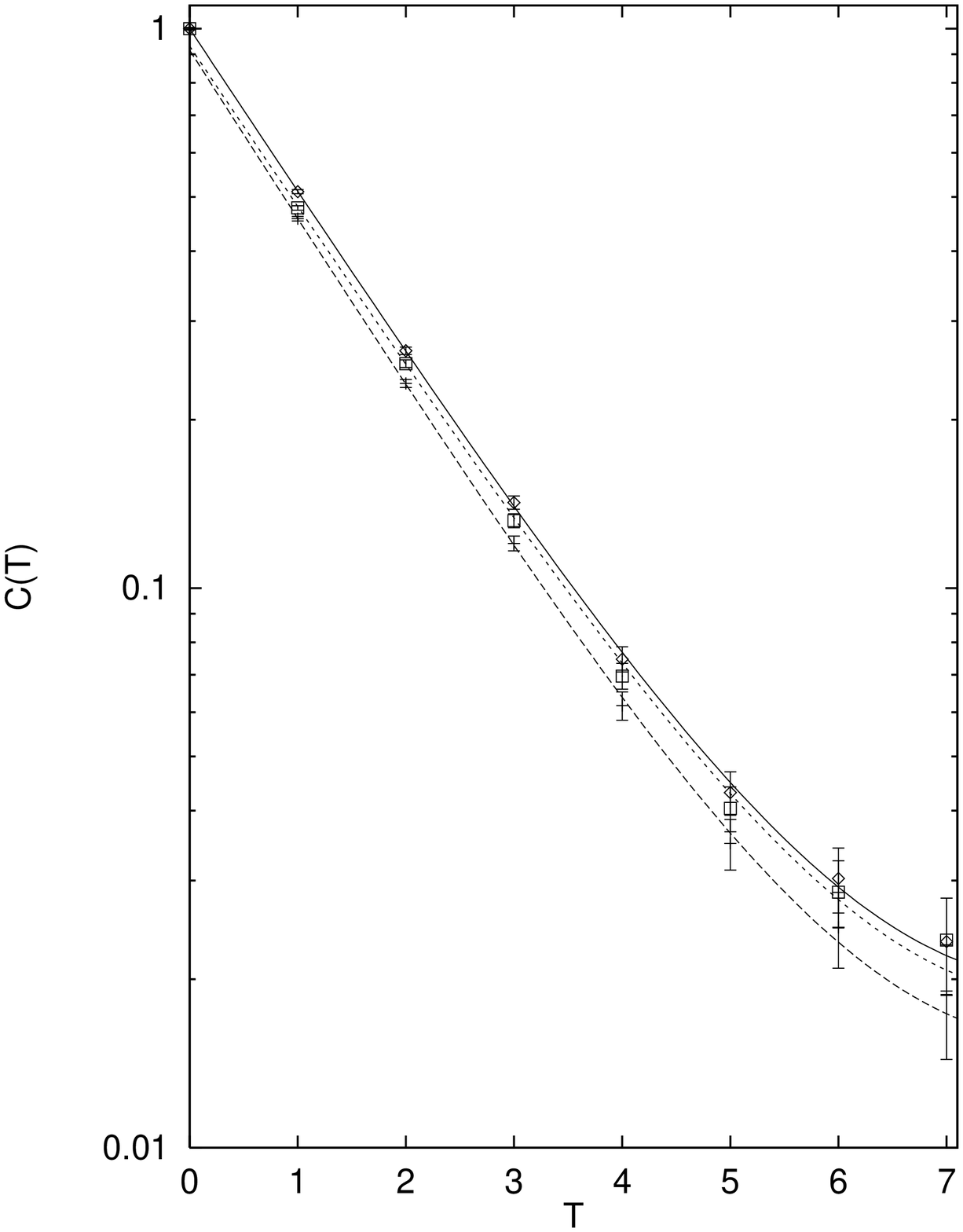}
        \vspace{-0.1cm}
	{\bf Figure 1}
        \end{center}
        \end{figure}

        \begin{figure}
        \begin{center}
        \leavevmode
        \epsfxsize=6.0in
        \epsfbox{ 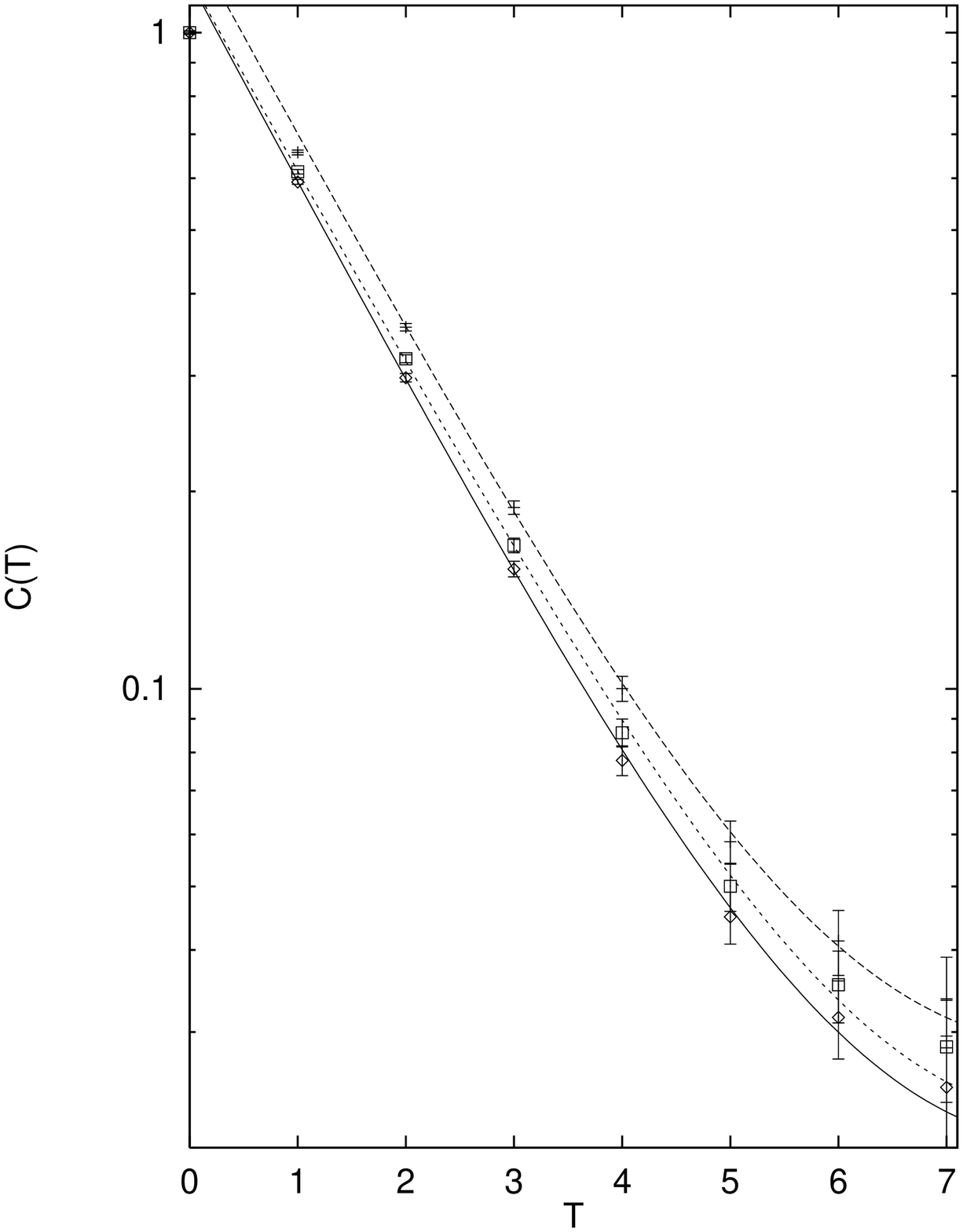}
        \vspace{-0.1cm}
	{\bf Figure 2}
        \end{center}

        \end{figure}

        \begin{figure}
        \begin{center}
        \leavevmode
        \epsfxsize=6.0in
       \epsfbox{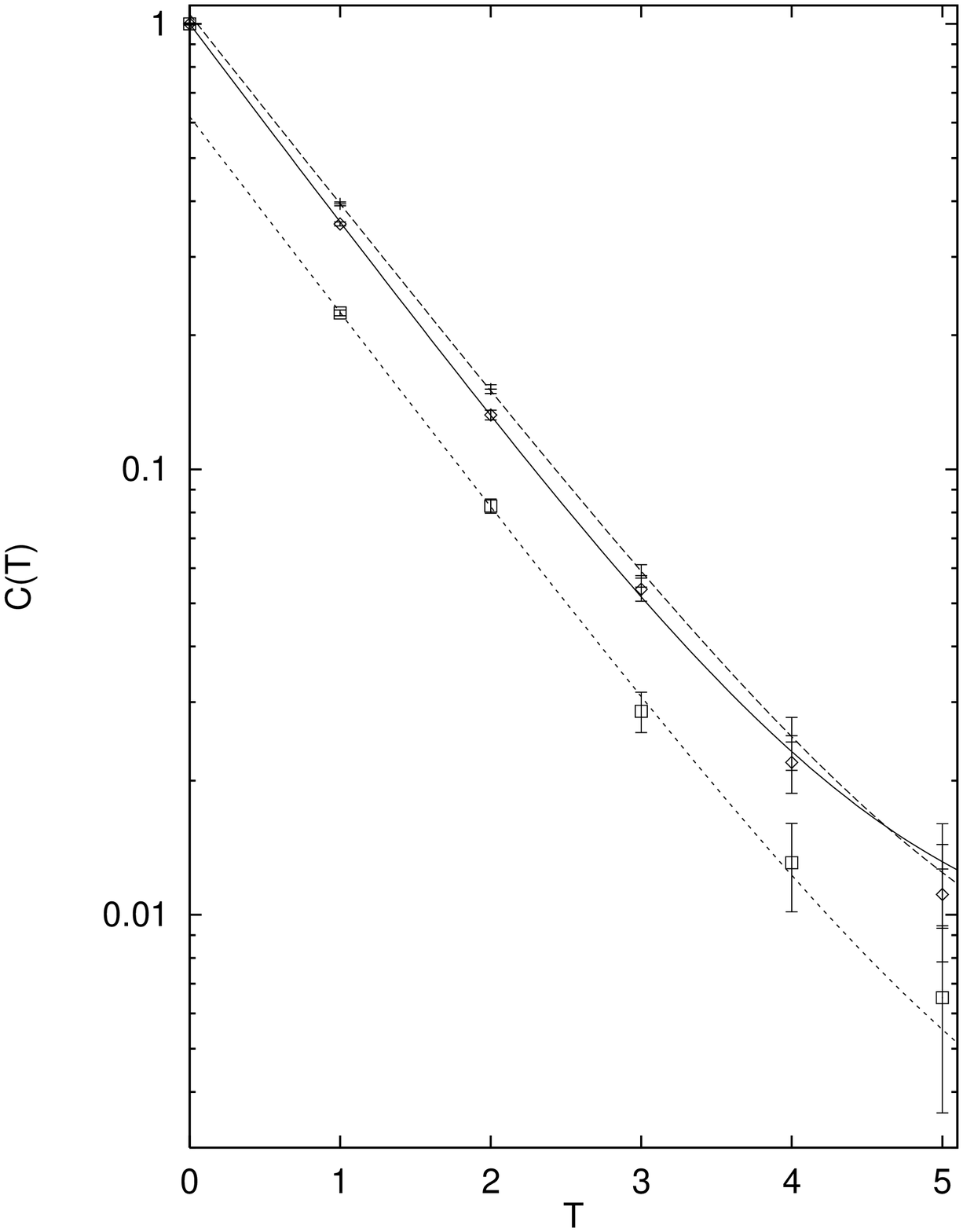}
        \vspace{-0.1cm}
	{\bf Figure 3}
        \end{center}

        \end{figure}

        \begin{figure}
        \begin{center}
        \leavevmode
        \epsfxsize=6.0in
       \epsfbox{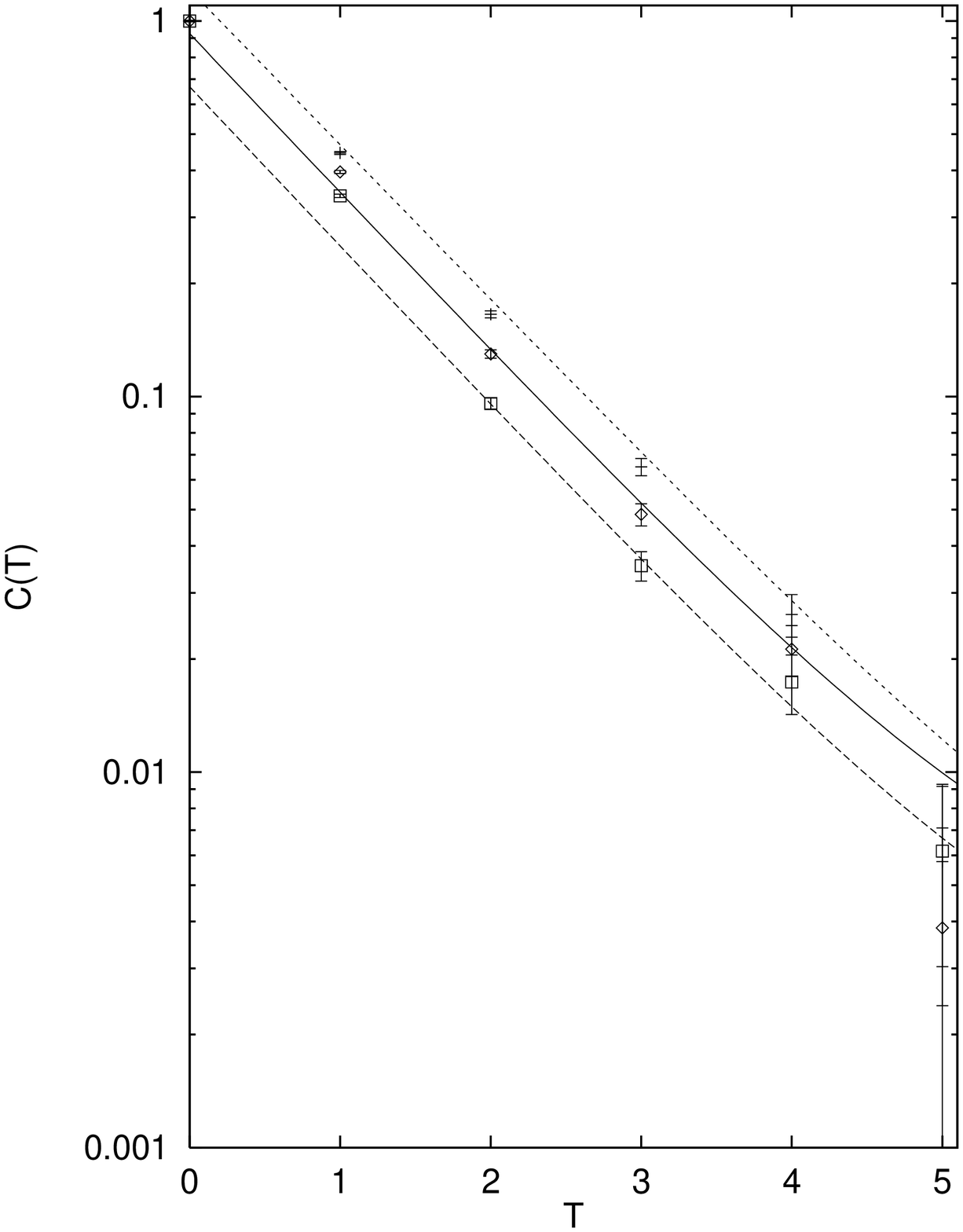}
        \vspace{-0.1cm}
	{\bf Figure 4}
        \end{center}

        \end{figure}


\begin{thebibliography}{99}

\bibitem{thoopolymon} G.\ 't Hooft, Nucl. Phys. {\bf B79},276(1974),
A.\ M.\ Polyakov, JETP Lett. {\bf 20},194(1974).

\bibitem{witten}  N. Seiberg and E. Witten Nucl. Phys. {\bf B426} 19 (1994).

\bibitem{thoo} G. 't Hooft Nucl. Phys. {\bf B190}[FS3] 455 (1982).


\bibitem{jssnrw} J. D. Stack, S. N. Neiman, and R. J. Wensley,
Phys. Rev. D {\bf 50} 3399 (1994).

\bibitem{bali} G. S. Bali, V. Bornyakov, M. Muller-Preussker, and 
K. Schilling  Phys Rev D {\bf 54} 2863 (1996).

\bibitem{polikarpov} See references in M.\ I.\ Polikarpov, 
Nucl. Phys. B (Proc. Suppl.) 53 (1997).

\bibitem{kron} A. S. Kronfeld, M. Laursen, G. Schierholz, and U.-J. Wiese,
Phys. Lett. B {\bf 198},516(1987).

\bibitem{degrand}T.\ A.\ DeGrand and D.\ Toussaint,
Phys. Rev. D {\bf 22},2478,(1980).

\bibitem{diracsh} P.\ A.\ M.\ Dirac, Phys. Rev. {\bf 74},817(1948).

\bibitem{stack_vol} J.\ D.\ Stack, R.\ J.\ Wensley, and S.\ D.\ Neiman,
Phys. Lett. B {\bf 385}, 261 (1996).
\bibitem{ishikawa} K. Ishikawa, G. Schierholz, and M. Teper
Z. Phys. C {19} 327 (1983).

\bibitem{teper_fuzz} M. Teper , Physics Letters B {\bf 183} 345 (1986).

\bibitem{ape} The Ape Collaboration, Physics Letters B {\bf 192} 163 (1987).

\bibitem{schierholz} 
F.\ Brandstaeter, A.\ S.\ Kronfeld,
and G.\ Schierholz, Nucl. Physics {\bf B345} 709 (1990).

\bibitem{creutz}  M.\ Creutz, Phys. Rev. D {\bf 21},2308(1980).

\bibitem{overr} M.\ Creutz, Phys. Rev. D {\bf 36},515(1987).

\bibitem{mandula} J. Mandula and M. Ogilvie, Phys. Lett. B {\bf 248},156(1990).

\bibitem{hart_gribov} A.\ Hart and M.\ Teper, Phys. Rev. D {\bf 55}, 3756 (1997).
\bibitem{michael-teper} C.\ Michael and M.\ Teper, 
Physics Letters B {\bf 199} 95 (1987).

\bibitem{js7-2} See for example, R.\ J.\ Wensley and J.\ D.\ Stack,
Phys. Rev. Lett. {\bf 63},1764(1989). 


\bibitem{shuryak} T.\ Sch\"{a}fer and E.\ V.\ Shuryak,
Phys. Rev. Lett. {\bf 75} 1707 (1995).

\bibitem{hart} A.\ Hart and M.\ Teper, Physics Letters B
{\bf 371}, 261 (1996).

\bibitem{chen} D.\ Chen, presentation at the International
Conference on Lattice Field Theory, Boulder Colorado, July 1998.
     
\end{thebibliography}
\end{document}